\documentclass[12pt,letterpaper]{article}
\usepackage{opex3}
\usepackage{amsmath}
\usepackage{url}
\usepackage[normalem]{ulem}
\usepackage{hyperref}
\input{My_def_Opex.def}
\begin{document}


\title{Communication and Measurement}


\author{James P. Gordon}
\address{Rumson, NJ}

\section*{Abstract} I discuss the process of measurement in the context of a communication system. The set-
ting is a transmitter which encodes some physical object and sends it off, a receiver which
measures some property of the transmitted physical object (TPO) in order to get some in-
formation, and some path between the transmitter and the receiver over which the TPO
is sent. The object of the game here is to characterize the TPO, either as an end in itself
(research), or to examine its potential for information transmittal (communication). If the
TPO is a 'small' object then quantum mechanics is needed to play this game. In the course
of this work, I hope to broaden the concept of measurement in quantum mechanics to in-
clude noisy measurements and incomplete measurements. I suggest that quantum density
operators are logically associated with the transmitter and some portion of the path, and
that quantum measurement operators are logically associated with the rest of the path and
the receiver. Communication involves the overlap of a density operator and a measurement
operator. That is, the fundamental conditional probability of the communication process is
given by the trace of the product of a density operator and a measurement operator. 
\newline


\section*{Preface\footnote{The preface was written by Mark Shtaif, School of Electrical Engineering, Tel Aviv University, Israel, shtaif@eng.tau.ac.il.}}
In late 2002 Jim Gordon handed me a first draft of this manuscript, proposing that I help him turn it into a paper.
In the 1960s, while working for Bell Laboratories, Jim wrote some of the very first papers on quantum communications, essentially giving birth to that important field. Initially, he was interested in exploring the ways in which Shannon's results regarding information capacity could be extended in order to account for the constraints imposed by quantum physics \cite{G61,G62,G64}. He was also interested in the theory of quantum measurement and published one of the first papers on the subject \cite{G66}. In the 1970s, however, Jim drifted away from quantum communications as his main interests shifted to other areas of research. It was only after his retirement from Bell Laboratories in the late 1990s that he re-engaged with quantum communications, returning to problems that he hadn't resolved in the first round. Jim had not been following the literature that had accumulated on the subject in the interim three decades, and after being exposed to some of it during our discussions, he decided not to proceed with the publication of this article. His reasoning was that some of the concepts that he introduced in it had become known over the years.

James P. Gordon died on June 21 2013.  In June of the following year,  I was asked to give a talk  about his contributions to quantum communications at a special symposium held  in his memory at the Conference on Lasers and Electro-Optics  in San Jose,  California \cite{SympCleo14}.  While reviewing the materials, I rediscovered this paper in my files and decided to present it at the symposium \cite{MarkCleo14}. I found that although some of the ideas may have been discussed elsewhere, Jim's paper  also contained important original content. Even more importantly, I realized that his unfinished manuscript offered a unique glance into the workings of a brilliant mind.  I also thought that his insightful text would be particularly useful to individuals with some background in optics and quantum physics, who were interested in building up their intuition regarding quantum measurement and communication without resorting to formal or abstract mathematics.  With the permission of Jim's family and colleagues, I decided to make this manuscript public.

Since the author was no longer alive, submission to a peer-reviewed journal did not seem reasonable; instead, posting the manuscript on arxiv seemed to be the most natural choice. What you will find in what follows is the last existing version of Jim's original text from 2006, retyped without any (intentional) changes, except for the correction of a few obvious typos. It is possible that some unintentional errors were introduced in the process of retyping, and hence the original PDF can be found at \url{http://www.eng.tau.ac.il/~shtaif/jpgcnm.pdf}. The posting of this retyped version on arxiv was necessary because arxiv functions as an archival journal, which means that the article can conveniently be cited in future work. Those who find typos in the retyped manuscript are encouraged to contact me at shtaif@eng.tau.ac.il, so that new corrected versions can be posted.

\section*{Introduction}
The subject of measurement has a long history. It describes how one sees things. All of the
problems that come up in this fundamental area have not yet been satisfactorily solved. In
particular, the role of the observer (e.g.,you) has never been satisfactorily integrated into the
quantum theory. Quantum mechanics arose because the classical theory could not account
for the results of laboratory experiments on 'small' objects like atoms, nor could it account
for the spectrum of black body radiation. In graduate school I learned that if an object
were prepared in a quantum state $|\psi\rangle$ and one made a measurement of some observable
(measurable quantity) $\mathbf A$, with possible values $[a]$, the result would be some particular value
$a$ with probability $|\lip a|\psi\rip|^2$. This prescription is in accord with the uncertainty principle. It
envisages a "pure" state of the object and an "exact" measurement of the observable $\mathbf A$,
and yet the measurement result is often not certain. This scenario fits in with our picture
of a communication system. The transmitter prepares the system in the state $|\psi\rip$ and the
receiver measures the observable $\mathbf A$ and gets the value $a$. In the real world, of course, the
transmitter often cannot put the system in a pure state, the receiver often cannot make an
exact measurement, and the transmission path often adds some uncertainty to the result. In
other words, the transmitter, the path, and the receiver can each add noise to the system.
The more general idea of simultaneous inexact measurements of non-commuting observables
was introduced in 1965. In theory, this typically involves coupling the received TPO to
an auxiliary system whose initial quantum state is known, followed by exact measurement
of a set of commuting observables of the combination. In the case of two non-commuting
observables it can also be thought of as a minimally invasive incomplete measurement of one
observable followed by an exact measurement of the other. In this work I will try to give a
logical picture of the communication process, and of how measurements fit into this picture.

\section*{The Model}
Figure 1 shows the components of our model communication system. It consists of a transmitter,
which sends out some TPO coded according to the transmitter knob setting $t$. The
TPO then follows the path to the receiver, during which voyage it may be corrupted by
some noise or be otherwise affected. The receiver makes a measurement of some property
of the TPO, getting a meter reading $r$. The ability of the system to transmit information is
governed by the conditional probability distribution $P_{RT}(r/t)$, defined as the probability
that the receiver's meter reading is $r$ given that the transmitter's knob setting is $t$. Subscripts
on the probability distributions indicate locations, while their arguments give values. The
knob settings and the meter readings may be multidimensional. They may, for example,
correspond to some set of observables of the TPO. Since the meter always reads some
value, the values of $P_{RT}(r/t)$ when summed over $r$ must equal unity.
\begin{figure}[t!]
\centering\includegraphics[width=0.9\columnwidth]{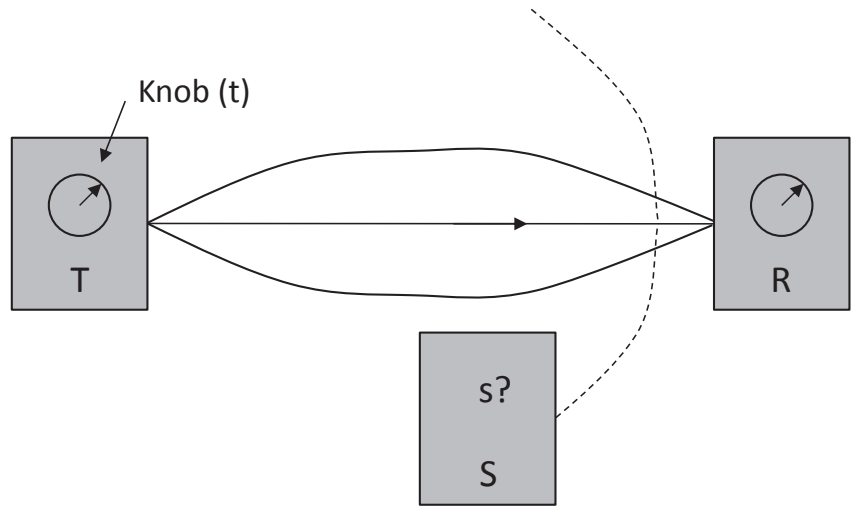}
\caption{Cartoon of the model communication system.}\label{PartialCoup}
\end{figure}
We want to know the properties of the TPO, as they are needed to design the best transmitter
and/or receiver. The problem then becomes how to describe the TPO somewhere
(anywhere) along the transmission path. To deal with this, we draw a surface $S$ cutting the
transmission path, and ask how the conditional probability distribution $P_{RT}(r/t)$ depends
on the properties of the TPO at $S$. This model covers most if not all laboratory experiments,
where the researcher wants to find something new about the nature of things. He/she builds
an apparatus which has knobs and meters, and works to try to describe what he/she is
looking at. He/she may try to minimize the random disturbances introduced by the transmission path.
Sometimes, however, such 'random' disturbances turn out to be important.
The model also applies to communication systems, where the engineer/scientist tries to de-
sign the transmitter and receiver in order to maximize the data rate, based on what he/she
thinks are the properties of the TPO and the transmission path. Most modern communi-
cation systems use electromagnetic fields sent through transmission lines or through space.
The TPO in this case might be the field received during some finite segment of time. In
the case of astronomy and similar endeavors, where the transmitter is beyond our control,
we can call it Nature, or God, as we like. In such a case, if we know something about the
properties of the path and the properties of the TPO, we can try to find out something
about the transmitter. Another source of corruption of a signal is inter-symbol interference
(ISI), where for example neighboring TPOs interfere with each other. I will not deal with
this problem, as it is a practical rather than a fundamental issue.

\section*{Analysis}
So how do we treat this communication problem? The classical answer envisions the possibility
of a complete and exact description of the TPO at $S$ by some set of real parameters $s$,
so that the conditional probability appears in the form
\bea P_{RT}(r/t)=\sum_s P_{RS}(r/s)P_{ST}(s/t). \label{10}\eea
Here, $P_{ST}(s/t)$ is the probability that the TPO at $S$ is in the state $s$ given that the transmitter
setting was $t$, while $P_{RS}(r/s)$ is the probability that the receiver records the value $r$ given
that the TPO was in the state $s$ at the location $S$. The sum over $r$ of $P_{RS}(r/s)$ and the
sum over $s$ of $P_{ST}(s/t)$ both equal unity, so that the sum over $r$ of $P_{RT}(r/t)$ is also unity.
Quantum mechanics was invented because this prescription often does not work.
Quantum mechanics envisions pairs of conjugate observables (measurable quantities), such
that it is not possible to determine a precise value for each at the same time. This is
Heisenberg's uncertainty principle. The canonical example is the position and momentum
of a single particle. TPO's are described by state vectors $|t\rip$ if they are in pure quantum
states, or more generally by density operators $\rho(t)$ whether they are in pure states or in
incoherent mixtures of pure states. Here $t$ represents some parameter set that specifies the
density operator. Density operators are Hermitian and non-negative, normalized to unit
trace. That is, density operators must satisfy the relation
\bea \mathrm{Trace}\big(\rho(t)\big)=1. \label{20}\eea
Representations of density operators are Hermitian matrices whose elements are formed using
the eigenstates of some complete set of commuting observables of the TPO. The diagonal
elements of any such representation are non-negative, sum to the trace of $\rho$ and therefore
to unity, and are generally acknowledged to form a probability distribution for the exact
measurement of those observables. It is of note that such eigenstates need not be possible
state vectors of the TPO. For example, the position $x$ of a single one-dimensional particle has
eigenstates $|x\rip$ which can be used to form the representation $\lip x|\rho|x'\rip$ of a particle in the state
$\rho$. The diagonal elements $\lip x|\rho|x\rip$ of this representation give the probability distribution for
the observable $x$ of the particle. The eigenstates of $x$ are not possible states of the particle
since their norm $\lip x|x\rip$ is infinite, but they do form a complete set since
$\int\df x|x\rip\lip x|+\mathbf I$, where
$\mathbf I$ is the identity operator. I will come back to this point later.

In our model communication system, the transmitter prepares the TPO, and sends it off on
the path to the receiver. It arrives at the location $S$ described by a density operator, which
we will label $\rho_{ST}(t)$, dependent on the transmitter setting $t$ and on the properties of the
path from the transmitter to the location $S$. It can have no dependence on the receiver or
on the part of the path from $S$ to the receiver. Since there are a number of different possible
representations of $\rho_{ST}(t)$, there is no unique set of values $s$ at $S$ as there is in the classical
picture. This is one of the important aspects of quantum mechanics.
The receiver is a measuring device. It measures some set of observables of the TPO, either
well or badly, and gets a result $r$ with the probability $P_{RT}(r/t)$. The most general way of
describing this process would seem to consist of a relation in the form
\bea P_{RT}(r/t) = \mathrm{Trace}\big(\sigma_{RS}(r)\rho_{ST}(t)\big), \label{30}\eea
where the operator $\sigma_{RS}(r)$ of Eq.(3) is a measurement operator. For example, the probability
distribution formed by the diagonal elements $\lip r|\rho|r\rip$ of any representation of the TPO can
be represented in the form of Eq (\ref{30}), with $\sigma(r) = |r\rip\lip r|$. In our communications milieu, this
would be the the result of an exact measurement of the set of commuting observables that
creates the representation. The measured value, $r$, would be the set of eigenvalues of that
set of observables for the eigenstate $|r\rip$.

I call the operator $\sigma_{RS}(r)$ of Eq.(\ref{30}) a measurement operator since it portrays the receiver's
role in the communication process. My thesis here is that measurement operators may be
given a broad scope, limited only by a few necessities. A measurement operator depends
only on the receiver's result $r$ and on the properties of the path from $S$ to the receiver. It
can have no dependence on the transmitter setting $t$ or on the part of the path from the
transmitter to $S$. The receiver may know or guess the properties of the TPO being sent, and
it may also know of any restrictions placed on the scope of the transmitter's choices, but it
knows nothing a priori about which of these choices has been made.

There are two basic requirements for any measurement operator. First, since $P_{RT}(r/t)$ must
be a non-negative real number for any possible density operator representing the TPO, it
follows that $\sigma_{RS}(r)$ must be Hermitian and non-negative, just as is a density operator. Second,
since $P_{RT}(r/t)$ is a probability distribution, summing over $r$ to unity, the measurement
operators must sum over $r$ to the identity operator. That is, the measurement operators
must satisfy the relation
\bea \sum_r\sigma(r) = \mathbf I.\label{40}\eea
The measurement operators must therefore form a complete set for the TPO. Their properties
need not be otherwise restricted. In our discussion of a 1D particle, the operator $|x\rip\lip x|$ is
an example of a measurement operator, even though it cannot be a density operator. More
precisely, it represents a limiting case of a real world measurement operator, since a truly
exact measurement of a continuous observable such as $x$ is beyond our capacity. The division
of quantum operators into two types, namely density operators which have unit trace and
measurement operators which satisfy a completeness relation, makes sense to me because
there are examples of each which do not fit into the other category.

In the early days of quantum mechanics, measurements were discussed only in the context of
representations of the density operator, thus implicitly involving only exact measurements
of complete sets of commuting observables. As I understand it, Heisenberg arrived at the
uncertainty principle by considering the measurement process, but the result was viewed only
through the properties of the density operator. More recently, with the advent of lightwave
communications, the scope of measurement operators was broadened to include inexact
simultaneous measurements of non-commuting observables, necessarily inexact because the
uncertainty principle forbids simultaneous exact measurements of such observables. This
type of measurement involves over-complete sets of measurement states, such as the coherent
states of the simple harmonic oscillator. Here I propose to take the measurement operator
concept one step further, that is, to broaden the scope of measurement operators to include
any operators $\sigma_{RS}(r)$ that satisfy the two basic requirements. It may not be easy or even
possible to imagine a device to realize every such measurement operator, but it should
be possible to devise a measurement operator to correspond to any imaginable device for
measuring any set of observables of a TPO. Some representative examples are discussed
below.

If the path is noisy, the density operators and measurement operators representing the TPO
will change for different positions of $S$ along the path, but the probability $P_{RT}(r/t)$ must re-
main independent of the location of $S$. This condition leads to some interesting relationships,
as we shall see later.

If the effects of the path are negligible, and the TPO at $S$ is actually in a pure quantum state
$|\psi\rip$ then $\rho_{ST}(\psi) = |\psi\rip\lip\psi|$. If in addition the receiver can accurately measure an observable
$\mathbf A$ whose eigenvalues are $[a]$, then $\sigma_{RS}(a) = |a\rip\lip a|$. Thus we come to my graduate school
result, namely that the probability of the measurement result a given the quantum state $|\psi\rip$
is
\bea P_{RT}(a/\psi)=\mathrm{Trace}\big(\sigma_{RS}(a)\rho_{ST}(\psi)\big) = \mathrm{Trace}\big(|a\rip\lip a||\psi\rip\lip\psi|\big) = |\lip a|\psi\rip|^2. \label{50}\eea
The description of the TPO at $S$ in terms of a quantum density operator automatically
satisfies the uncertainty principle with respect to the ability of the transmitter to ascribe
values to the observables of the TPO. Likewise, the description of the measurement process
at $S$ in terms of a quantum measurement operator automatically satisfies the uncertainty
principle with respect to the ability of the receiver to measure these values. The point
is that neither the transmitter nor the receiver can ever know more about the TPO than
the uncertainty principle permits. In the communication process, the uncertainty principle
applies independently to the preparation and to the measurement of any TPO.

There is considerable overlap in the realms of density operators and measurement operators.
There are differences, however. For example, measurements of only one of a non-commuting
conjugate pair of observables correspond to measurement operators whose traces are infinite,
so they cannot serve as density operators.

This completes our brief formal analysis. Density operators are discussed in most texts
on quantum mechanics. In the rest of this paper, I will discuss various examples to give
substance to the idea of measurement operators. For simplicity, I will deal with TPO's in
the form of simple harmonic oscillators (SHO)s, and two level systems (TLS)s. Linear fiber
optical communication systems can be modelled by SHOs transmitted at a rate given by the
system bandwidth. In the modern parlance of quantum computation, TLSs are examples of
qbits. Thus there is some substance to this work. Bold face will be used for operators, but
not for values. The context can also help to distinguish values from operators.

\section*{The simple Harmonic Oscillator}
The classical picture of the SHO envisions the possibility of exact prescription and of exact
measurement of both the position and momentum of a particle in a harmonic potential well.
From the standpoint of communications, this implies that the information transfer possible
using a single SHO of finite energy is limited only by noise. As we know, this classical picture
runs into fatal trouble trying to explain things such as the law of black body radiation. The
quantum picture of the SHO is more complicated, and is really quite counter-intuitive. The
conjugate variables position and momentum are represented by non-commuting operators.
The result is a ladder of energy states separated by the quantum of energy, the lowest of which
has an energy one-half quantum above the bottom of the potential well. Either position or
momentum can still be prescribed and/or measured as exactly as one pleases, but only at
the expense of greater and greater uncertainty, and therefore greater expected energy, in the
conjugate variable. Communication, with an energy constraint, is thereby limited even in
the absence of noise. Elements of the quantum theory of the SHO are reviewed in Appendix
\ref{App1}.

The SHO provides a fertile ground for studying the communications problem. The transmitter
may be able to send SHOs in either energy states or minimum uncertainty 'coherent'
states. The path may involve either loss or gain, or both. The receiver may measure energy,
or position or momentum, or both. I will examine a number of these possibilities.

First, let us look at the case where the SHOs are encoded and measured by energy. Suppose
that the transmitter can emit SHO's with exactly $N$ energy quanta. Suppose also that
the path is purely lossy, and that the receiver can measure the exact number of quanta
in the received SHO. The probability that the receiver measures $M$ quanta given that the
transmitter sent $N$ quanta is given by the binomial distribution
\bea P_{RT}(M/N) = C_M^N(p_{RT})^M(1-p_{RT})^{N-M} \equiv B_M^N(p_{RT}),\label{60}\eea
where $C_M^N$ is the binomial coefficient, while $p_{RT}$ is the probability that a quantum will survive
the loss in the path between the transmitter and the receiver. The last expression defines
the binomial form $B_M^N(p_{RT})$.

The energy states $|n\rip$ are labelled by the number of quanta, and comprise a complete set,
so they can also be used as measurement states. They satisfy the relations $\lip n|m\rip = \delta(n,m)$
and $\sum |n\rip\lip n| = \mathbf I$.

If we put the surface $S$ immediately in front of the receiver, the density operator representing
the transmitted SHO will reflect the binomial probability distribution, while the measurement
operator will reflect the exact number of received quanta. It is pretty clear that the
density and measurement operators in this example are given by
\bea \rho_{ST}(N) = \sum_n B_n^N(p_{RT})|n\rip\lip n|\hspace{2cm}\sigma_{RS}(M) = |M\rip\lip M|,\label{70}\eea
since $B_n^N(p_{RT})$ is the probability the n quanta reached the receiver given that $N$ quanta
were sent. In this case, the conditional probability $P_{RT}(M/N)$ of Eq.(\ref{60}) can be written as
$\lip M|\rho_{ST}(N)|M\rip$, which is one of the diagonal elements of the density operator written in the
number representation. This accords with the standard recipe for measurements in quantum
mechanics. However, in our extended view of measurement operators, we can equally well put
the surface $S$ immediately after the transmitter. In this case the density operator for the sent
SHO represents a pure state of $N$ quanta, and so the binomial distribution must therefore be
represented in the measurement operator. The density and measurement operators transform
to
\bea \rho_{ST}(N) = |N\rip\lip N|\hspace{2cm}\sigma_{RS}(M) = \sum_n B_M^n(p_{RT})|n\rip\lip n|.\label{80}\eea
This relation is an example of an inexact measurement operator. It follows since $P_{RT}(M/N)$
must be independent of the location of the surface $S$, and it makes sense since $B_M^n(p_{RT})$ is
the probability that $M$ quanta reached the receiver given that $n$ quanta were sent.

In more generality, we can locate $S$ anywhere between transmitter and receiver. In this case
Eqs.(\ref{70}) and (\ref{80}) change to
\bea \rho_{ST}(N) = \sum_n B_n^N(p_{ST})|n\rip\lip n|\hspace{2cm} \sigma_{RS}(M) = \sum_n B_M^n(p_{RS})|n\rip\lip n|,\label{90}\eea
where $p_{RS}p_{ST}=p_{RT}$. In these three examples, the prerequisites are satisfied, as they must
be. That is, in each case $\mathrm{Trace}(\rho)$ is unity, the sum of $\sigma$ over the measurement results gives the identity operator, and $\mathrm{Trace}(\sigma\rho)$ gives the conditional probability of the measurement
result, given the transmitter setting. For example, from Eq.(\ref{90}) we have
\bea P_{RT}(M/N) = \mathrm{Trace}\big(\sigma_{RS}(M)\rho_{ST}(N)\big)=\sum_nB_M^n(p_{RS})B_n^N(p_{ST})=B_M^N(p_{RT}),\label{100}\eea
in accord with Eq. (\ref{60}). Because all of the density and measurement operators are diagonal
in the same (here energy) representation, these results have the form of the classical Eq.(\ref{10}),
with $s = n$, $t = N$, and $r = M$. The role of quantum mechanics is the quantization
of the energy. This example however illustrates an important communications limitation
imposed by quantum mechanics. Communication using this system is maximized if there is
no attenuation in the path. In this case Eq.(\ref{60}) reduces to $P_{RT}(M/N) = \delta(M,N)$. If the
system is only allowed to use SHO's with energies no greater than $N$ quanta, the information
transfer per SHO is limited to $\log_2(N)$ bits. There is no way of encoding the SHO's that
exceeds this limit under the same constraint.

As I noted above, the measurement operators are independent of the properties of the transmitter
and the portion of the path prior to $S$. Thus for any $\rho_{ST}(t)$, the receiver represented by
Eq.(\ref{90}) would measure $M$ quanta with probability $P_{RT}(M/N) = \sum_n B_M^n(p_{RS})\lip n|\rho_{ST}(t)|n\rip$.
The measurement operators of Eqs. (\ref{80}) and (\ref{90}) describe inexact measurements of the number of
quanta at $S$, inexact because of the attenuation in the path from $S$ to the receiver.

The basic properties of the distribution $B_M^n(p)$ as a function of $n$ are
\bea \sum_n B_M^n(p) = \frac 1 p\hspace{0.5cm}n_{max}\simeq M/p-0.5\hspace{0.5cm}\Delta n\simeq \frac{2.355}{p}\sqrt{M(1-p)+p/2},\label{110}\eea
where $p<1$, $n_{max}$ is the location of the peak of the distribution, and $\Delta n$ is its full width at
half maximum (fwhm), based on the curvature of the peak along with a Gaussian approximation to its shape. The actual distribution is somewhat skewed toward higher values of $n$, but the value of the fwhm is reasonably accurate. For example, if $M = 20$ and $p = 0.5$, the
distribution peak is approximately at $n= 39.5$ (the values at $n = 39$ and $n = 40$ are equal),
and the fwhm is approximately 15 (the values nearest to the half maxima are at $n = 33$ and
$n = 48$).

An alternate regime for communication using SHO's is a coherent system, where phase
information as well as amplitude information is used. In quantum mechanics, the simplest
such system involves a transmitter emitting SHO's in minimum uncertainty coherent states,
and a receiver using coherent states as measurement states. As discussed in Appendix \ref{App1},
the coherent states are versions of the ground state displaced to other places in the phase
space of the SHO. When subject to pure attenuation, it is well known that coherent states
remain coherent states as their mean values (signals) decay. When subject to pure gain, it
is also well known that enough noise is added to maintain the validity of the uncertainty
principle.

Let us see how this situation plays out in our picture of measurements. The density operator
corresponding to the pure coherent state $\alpha\rip$ is denoted $|\alpha\rip\lip\alpha|$, where $\alpha$ is a complex number
which may be thought of as a classical signal. The measurement operator corresponding to
the pure coherent state $|\beta\rip$ is denoted $\pi^{-1}|\beta\rip\lip\beta|$, where $\beta$ is also a complex number. The
difference is due to the normalization required of the two operators. If the transmitter sends
the SHO in the state $|\alpha\rip$, and the path is loss free, the probability that an ideal coherent
receiver would record the value $\beta$ is given by $P_{RT}(\beta/\alpha) = \pi^{-1}|\lip\beta|\alpha\rip|^2 = \pi^{-1}\exp(-|\beta-\alpha|^2)$.
Thus, with both ideal preparation and ideal measurement in a coherent system, there is
uncertainty in the result, having a two dimensional Gaussian distribution that corresponds
to an effective single quantum of Gaussian noise.

To see this more clearly, we can generalize the above result. Suppose that the path adds
some extra Gaussian noise, and that $S$ is immediately in front of the receiver. The density
operator corresponding to signal plus Gaussian noise using the coherent state expansion is
given in Appendix \ref{App2}, Eq.(\ref{B11}). The same ideal coherent receiver would record the value �
with the probability
\bea P_{RT}(\beta/\alpha)&=&\frac{1}{\pi^2\bar n}\int\int \df^{(2)}(\xi)\exp\left(\frac{|\xi-\alpha|^2}{\bar n}\right)|\lip\beta|\xi\rip|^2\nonumber\\
& =& \frac{1}{\pi(\bar n+1)}\exp\left(-\frac{|\beta-\alpha|^2}{\bar n+1}\right),\label{120}\eea
where $\bar n$ is the mean number of noise quanta, and $|\xi\rip$ represents a coherent state.

For the same conditions, let us put $S$ immediately after the transmitter. Then $\rho_{ST}(\alpha) =
|\alpha\rip\lip\alpha|$ and we need a measurement operator which will reproduce the result of Eq(\ref{120}). The
answer is
\bea \sigma_{RS}(\beta) = \frac{1}{\pi^2\bar n}\int\int \df^{(2)}(\xi)\exp\left(\frac{|\xi-\beta|^2}{\bar n}\right)|\xi\rip\lip\xi|.\label{130}\eea
Thus, akin to the case of a lossy line with photon numbers being prescribed and measured,
a noisy coherent signal and a pure coherent measurement is equivalent to a pure coherent
signal and a noisy coherent measurement. One can see from Eq.(\ref{120}) that this communication
process involves an effective minimum of one quantum of Gaussian noise.

In the interest of simplicity, I am not taking into account whatever phase shifts may occur
in the path.

Other cases of interest are where the path involves loss or gain. First, consider the case of
pure loss, with $S$ immediately in front of the receiver. If the transmitter sends a SHO in the
coherent state $|\alpha\rip$, it will reach the receiver in the coherent state $|\alpha/\sqrt{L}\rip$, where $L\geq1$ is the
loss. If the receiver then measures the value $\beta$ corresponding to the coherent measurement
state $|\beta\rip$, we get for the conditional probability the result
\bea P_{RT}(\beta/\alpha) = \pi^{-1}|\lip\beta|\alpha/\sqrt{L}\rip|^2=\pi^{-1}\exp\left(-|\beta-\alpha/\sqrt{L}|^2\right).\label{140}\eea
This is straightforward, but now what if $S$ is immediately in front of the transmitter? The
density operator for the transmitted signal is then just $|\alpha\rip\lip\alpha|$, and we need to discover the
measurement operator. If we rewrite Eq (\ref{140}) as
\bea P_{RT}(\beta/\alpha) = \pi^{-1}\exp\left(-\frac{|\sqrt{L}\beta-\alpha|^2}{L}\right),\label{150}\eea
and compare it with Eqs.(\ref{120}) and (\ref{130}), the answer emerges as
\bea \sigma_{RS}(\beta) = \frac{1}{\pi^2}\frac{L}{L-1}\int\int\df^{(2)}(\xi)\exp\left(\frac{|\xi-\sqrt{L}\beta|^2}{L-1}\right)|\xi\rip\lip\xi|.\label{160}\eea
Thus it happens again that an attenuated signal and a pure measurement is equivalent again
to a pure signal and a noisy measurement. This is not too surprising, but I was happy to
find this result.

The case where the path has pure gain turns out to be the inverse of the case of pure loss. To
see this, suppose that $S$ is first placed at the transmitter. Then the density operator for the
transmitted SHO is simply $\rho_{ST}|\alpha\rip\lip\alpha|$. The best we can do in coherent measurements is
to use coherent measurement states. So suppose that the measurement operator is $\sigma_{RS}(\beta) = (\pi G)^{-1}|\beta/\sqrt{G}\rip\lip\beta/\sqrt{G}|$. (Remember that the value $\beta$ is what the receiver records, and that the
path has gain $G$.) In this case the conditional probability evaluates to
\bea P_{RT}(\beta/\alpha) = \frac{1}{\pi G}\exp\left(|\alpha-\beta/\sqrt{G}|^2\right).\label{170}\eea
When $S$ is moved to the receiver, the measurement operator becomes $\sigma_{RS}(\beta) =\pi^{-1}|\beta\rip\lip\beta|$, and
the density operator necessary to preserve the conditional probability Eq(\ref{170}) is
\bea \rho_{ST}(\alpha) = \frac{1}{\pi(G-1)}\int\int\df^{(2)}(\xi)\exp\left(\frac{|\xi-\sqrt{G}\alpha|^2}{G-1}\right)|\xi\rip\lip\xi|.\label{180}\eea
This density operator has the form of signal plus noise, with $\bar n = G - 1$. This is not too
surprising, but it is impressive that the uncertainty principle (which is responsible for the
coherent states) also demands the emission of the noise that is well known to accompany
gain in a transmission system.

As we have just seen, in the coherent state picture the cases of pure loss and pure gain
are intimately related. Normalization aside, the density operator for pure gain with $S$ at
the receiver has the same form as the measurement operator for pure loss with $S$ at the
transmitter. These are given in Eqs.(\ref{180}) and (\ref{160}). The question arises (courtesy of M.S.),
does the same apply to the case where number states are used. His answer is yes. From
the measurement operator in Eq.(\ref{80}), one may guess that the density operator for the case
of pure gain with $S$ at the receiver, when $N$ photons are sent, would be
\bea\rho_{ST}(N) = G^{-1}\sum_n B_N^n(G^{-1})|n\rip\lip n|, \label{190}\eea
and this is the correct answer. For example, the mean number of photons predicted by this
distribution is $GN + G - 1$, which represents the input number multiplied by $G$ plus the
$G - 1$ photons expected from the spontaneous emission that accompanies the gain.

Since the density operators and measurement operators are independent entities, there are
many combinations that can be examined. One that I find pedagogically interesting is to
transmit a number state, and measure it using coherent states, with no loss or gain in the
path. The result is
\bea P_{RT}(\beta/n)=\pi^{-1}|\lip\beta|n\rip|^2 = \frac{1}{\pi n!}(\beta\beta^*)^n\exp(-\beta\beta^*).\label{200}\eea
This probability distribution peaks at $|\beta|^2 = n$ independent of the phase of $\beta$. It is a
smoothed measure of how the number states are distributed in the phase space of the SHO.

\subsection*{The Wigner Picture}
To visualize these results it is useful to appeal to the Wigner distributions. As discussed
in Appendix \ref{App2}, these are quasi-classical distributions in the phase space of the SHO that
are isomorphic with the quantum density operators. They differ from classical probability
distributions in that for most quantum states, the corresponding Wigner distributions have
negative values in various regions of the phase space. For the cases of signal plus Gaussian
noise discussed above, however, they are positive definite, and lead to a classical picture of
the communications process. Furthermore, they can give a classical description of quadrature
squeezed states. I think that it is fair to think of any SHO whose Wigner distribution is
positive definite as being in a 'classical' state.

There are three important features of the Wigner picture. First, as just mentioned, Wigner
distributions are isomorphic with quantum density operators, so they can accurately
represent quantum states. Second, as also noted in appendix \ref{App2}, the integral over the phase
space of the SHO of the product of two Wigner distributions is proportional to the trace
of the product of the corresponding two density operators. With a change in the normalization,
Wigner distributions can also represent measurement operators. Thus, the crucial
conditional probability $P_{RT}(r/t)$ of the communications process can be correctly viewed as
the overlap of two Wigner distributions in the phase space, in the form of Eq.(\ref{10}). Third, of
considerable importance to the photonic communications business, the Wigner distribution
for the SHO obeys the classical equations of motion, with no quantum corrections. This is
true also for linear couplings of the SHO to heat baths, yielding attenuation or gain. For a
particle in an anharmonic potential, the Wigner distribution obeys the classical equations of
motion to second order in the size of the energy quantum. As a result, the first and second
moments of Wigner distributions always behave classically. One must go to third and higher
moments to find quantum corrections. Although as far as I know it has not yet been proven,
I do not think it is much of a stretch to presume that for photonic propagation in glass fibers,
with their weak non-linearity, the Wigner picture yields a very close classical approximation
to the exact quantum picture. It does so if an adequate description of the field in the fiber
consists of signal plus Gaussian noise, which depends only on first and second moments. This is,
for example, why there are no significant quantum corrections to a classical description
of soliton propagation in fibers using the Wigner picture.

In essence, the Wigner picture treats the half quantum of zero point field of the SHO as
an integral part of a total classical Gaussian noise field. In particular, the zero point field
suffers attenuation and/or gain just as do any additional noise fields. This has implications
in regard to noise generation from media that attenuate or amplify the field. Attenuators
must be noise generators in order to maintain the zero-point field. As it turns out, the
spontaneous emission of noise in the Wigner picture is the same for both amplifying and
attenuating media. We will see later how this may be rationalized.

Let us look for the Wigner picture of measurement states. The Wigner picture uses the
coordinates $(p,q)$ of phase space. If we rewrite the coherent state $|\alpha\rip$ as $|p,q\rip$, where $\alpha=(1/\sqrt{2})(q+ip)$, the completeness relation becomes $(1/\sqrt{2})\int\df p\int\df q|p,q\rip\lip p,q| = \mathbf I$. Thus, if we use the coordinates of phase space, we find that
\bea \sigma(p,q) = (1/\sqrt{2})\rho(p,q).\label{210}\eea
It follows from Eq. (\ref{50}) of Appendix {\ref{App2} that the Wigner distribution corresponding to a measurement state is
\bea W(p,q/\sigma) = \int\df q'\lip q-\frac 1 2 q'|\sigma|q+\frac 1 2 q'\rip\exp(iq'p). \label{220}\eea
This form works for all measurement operators, including those with in�nite trace. For
example, the Wigner distribution corresponding to the measurement operator $|q''\rip\lip q''|$ is $W(p,q)=\delta(q-q'')$. This Wigner distribution is not integrable because it is independent of
p, but it is nonetheless applicable to the communications problem, since its overlap with the
Wigner distribution corresponding to any density operator is finite.

In view of Eq. (\ref{210}) and Eq. (\ref{B6}) of Appendix \ref{App2}, it follows that the basic conditional probability
of the communication process can be written in the classical form
\bea {P_{RT}}(r/t) = \int\df p\int\df q W_{RS}(p,q/r)W_{ST}(p,q/t),\label{230}\eea
where $W_{ST}$ is the Wigner distribution corresponding to the density operator $\rho_{ST}$ and $W_{RS}$ is the Wigner distribution corresponding to the measurement operator $\sigma_{RS}$.

As given in appendix \ref{App2}, the basic Wigner distribution for the SHO corresponding to signal
plus Gaussian noise is given by
\bea\label{240} W(p,q/p',q') = \frac{1}{\pi(2\bar n+1)}\exp\left(-\frac{(q-q')^2+(p-p')^2}{2\bar n+1}\right),\eea
where $\bar n$ is the mean number of noise quanta. This is a two dimensional Gaussian centered
on the signal $(p',q')$. Note the extra half quantum of zero-point noise represented in the
quantity $2\bar n+1$, where $\bar n$ is the mean number of noise quanta. The case $\bar n = 0$ is the Wigner
distribution for a coherent state, which consists of a signal plus half a quantum of Gaussian
noise.

Consider then the cases of coherent communication discussed above. With a lossy path, and
$S$ at the receiver, we have
\bea W_{ST}(p_S,q_S/p_T,q_T) = \pi^{-1}\exp\big(-(q_S-q_T/\sqrt{L})^2 - (p_S-p_T/\sqrt{L})^2\big)\label{250}\eea
and
\bea W_{RS}(p_R,q_R/p_S,q_S) = \pi^{-1}\exp\big(-(q_R-q_S)^2 - (p_R-p_S)^2\big),\label{260}\eea
which yields from Eq. (\ref{230})
\bea P_{RT}(p_R,q_R/p_T,q_T) = \frac {1}{2\pi}\exp\left(-\frac{(q_R-q_T/\sqrt{L})^2 + (p_R-p_T/\sqrt{L})^2}{2}\right)\label{270}.\eea
In this communication there is an effective one quantum of noise, half of which is associated
with the transmitted SHO, and half is associated with the measurement. With the same
lossy path, and $S$ at the transmitter, we get
\bea W_{ST}(p_S,q_S/p_T,q_T) = \pi^{-1}\exp\big(-(q_S-q_T)^2 - (p_S-p_T)^2\big)\label{280}\eea
and
\bea W_{RS}(p_R,q_R/p_S,q_S) = \frac{L}{\pi(2L-1)}\exp\left(-\frac{(q_R\sqrt{L}-q_S)^2+(p_R\sqrt{L}-p_S)^2}{2L-1}\right),\label{290}\eea
which again leads to Eq. (\ref{270}) via Eq. (\ref{230}). Looking back up the path, the measurement
becomes noisy in order to maintain the same signal to noise ratio. Similar results apply to
the case where the path has gain.

States of the SHO which are quadrature squeezed versions of signal plus additive Gaussian
noise also have positive definite Wigner densities, and so we can think of them as classical
states. In recent years there has been much interest in such squeezed states of the radiation
field, and while they have proven difficult to make, some squeezing has been achieved. Most
attention has been focused on squeezed versions of the vacuum state. A substantial problem
is that the squeezed vacuum reverts to the zero point field when it is attenuated. In the
Wigner picture, this is simply because the initial squeezed field is attenuated, while the zero
point field grows to replace it.

The Wigner density for the quadrature squeezed vacuum has the form
\bea W(p,q) = \pi^{-1}\exp\big(-\eta q^2-p^2/\eta\big),\label{300}\eea
where $\eta$ is a positive real parameter. If $\eta$ is greater than one, the state is squeezed in the
$q$ direction and enlarged in the $p$ direction, while the reverse is true if $\eta$ is less than one.
All squeezed vacuum states take up the same area in the phase space, as required by the
uncertainty principle. The propagation of a squeezed vacuum state in a lossy path can be
determined from the foregoing. If a transmitter emits a SHO in a squeezed vacuum state, and
an ideal coherent detector lies at the end of the path, then if we locate $S$ at the transmitter,
we have
\bea W_{ST}(p_S,q_S/t) = \pi^{-1}\exp\big(-\eta q_S^2-p_S^2/\eta\big)\label{310}.\eea
Because the state preparation and the state measurement are independent processes, the
measurement is again described by the Wigner density of Eq. (\ref{290}). Thus we get, according
to Eq. (\ref{230}),
\bea P_{RT}(p_R,q_R/t) &=& \frac{L}{\pi\sqrt{(2L-1+\eta^{-1})(2L-1+\eta)}}\times\nonumber\\
&&\exp\left(-\frac{Lq_R^2}{2L-1+\eta^{-1}}-\frac{Lp_R^2}{2L-1+\eta}\right)\label{320}.\eea
If we now move $S$ to the receiver, where the measurement is described by Eq. (\ref{260}), in order
to preserve Eq. (\ref{320}) the Wigner density of the transmitted state evaluates to
\bea W_{ST}(p_S,q_S/t) &=& \frac{L}{\pi\sqrt{(L-1+\eta^{-1})(L-1+\eta)}}\times\nonumber\\
&&\exp\left(-\frac{Lq_S^2}{L-1+\eta^{-1}}-\frac{Lp_S^2}{L-1+\eta}\right)\label{330}
.\eea
One can see how the field transforms from the initial squeezed state at $L = 1$ to the zero
point field when $L\gg1$. The field described by Eq. (\ref{330}) can be decomposed into the sum
of two fields, one, the incident squeezed field, decaying with $p,q\propto\sqrt{1/L}$
and the other, the zero point field, growing as a result of the combination of emission and absorption, with
$p,q\propto\sqrt{1-1/L}$.

\subsection*{Incomplete measurements}
So far this discussion of the SHO has been concerned with what we may call complete
measurements, which may be de�ned as those measurements that leave no residual information
about the transmitter setting. There is another class of measurements, incomplete
measurements, which can leave such information. A canonical example of an incomplete
measurement is given by the measurement operator
\bea \sigma_{RS}(q_R)  &=&  \sqrt{\frac\eta\pi}\exp\big(-(\eta(q_R-\mathbf q)^2\big)\nonumber\\ &=& \sqrt{\frac\eta\pi}\int\df q\exp\big(-(\eta(q_R- q)^2\big)
|q\rip\lip q|,\label{340}\eea
where the second form has been expanded in the position representation. The result of such
a measurement is
\bea P_{RT}(q_R/t)  &=&  \mathrm{Trace}\big(\sigma_{RS}(q_R)\rho_ST(t)\big)\nonumber\\  &=&  \sqrt{\frac\eta\pi}\int\df q\exp\big(-(\eta(q_R- q)^2\big)
\lip q|\rho_{ST}|q\rip.\label{350}\eea
This result bespeaks a measurement of the position variable $\lip q|\rho_{ST}|q\rip$ of $\rho_{ST}$ with an
accuracy controlled by the value of the real positive parameter $\eta$. Some information about
the transmitted SHO may still be available after such a measurement.

Some discussion is merited here. Since the measurement operator of Eq. (\ref{340}) has an infinite
trace, it cannot be renormalized into a density operator. For any real world measuring device,
however, some limitation on the momentum variable $p$ is always present, so one may argue
that the trace of any real world measurement operator must be finite. This is surely so, but I
believe that it leads only to analytical complexity. Also in the real world, the TPO after the
measurement may be left in a variety of states, depending on the measuring device. It may
be consumed, or not. There is a class of measurements, appropriately called non-demolition
measurements, which do not destroy the TPO. After any non-demolition measurement, the
density operator representing the TPO must reflect the result of the measurement $r$ as well
as the original transmitter setting $t$. Hence it must change. I denote a minimally invasive
measurement giving the result $r$ as one that leaves the TPO in the state
\bea \rho_{RT}(t,r) = \big(P_{RT}(r/t)\big)^{-1}\sqrt{\sigma_{RS}}\rho_{ST}\sqrt{\sigma_{RS}}\label{360}.\eea
In this relation, $\rho_{RT}(t,r)$ is the residual density operator, which now depends on both the
transmitter setting $t$ and the measurement result $r$. Since any measurement operator $\sigma_{RS}(r)$
is Hermitian and non-negative, it has a Hermitian and non-negative square root, which is
what is meant in Eq. (\ref{360}). To find this square root, one can diagonalize the measurement
operator and then take the positive square root of each diagonal element. The relation (\ref{360})
is the simplest way of creating a new density operator dependent on the original density
operator and on the measurement operator. It is a generalization of previous formulations
in that the measurement operator is a generalization of previous work. Note that if the
measurement is complete, the TPO is left in a state which depends only on the measurement
result. To take a simple example, if the TPO is a SHO, an ideal coherent measurement has
the measurement operator $\pi^{-1}|\beta\rip\lip\beta|$ whose square root is $\pi^{-1/2}|\beta\rip\lip\beta|$. According to Eq. (\ref{360}) a
SHO may at best be left, after the measurement, in the coherent state $|\beta\rip\lip\beta|$, which depends
only on the measurement result, and not at all on the transmitter setting.

If some information about the transmitter setting is left in the new density operator, it may
be harvested by a subsequent measurement. Suppose that we have two sequential measuring
devices $R_1$ and $R_2$. A minimally invasive measurement by $R_1$ giving a result $r_1$ leaves a TPO
in a state described by
\bea \rho_{R_1 T}(t,r_1) = \big(P_{R_1 T}(r_1/t)\big)^{-1}\sqrt{\sigma_{R_1S}}\rho_{ST}\sqrt{\sigma_{R_1S}}\label{370}.\eea
A subsequent measurement on the same TPO having the measurement operator $\sigma_{R_2R_1}(r2)$
gives the result
\bea P_{R_2R_1}(r_2/t,r_1) = \mathrm{Trace}\big(\sigma_{R_2R_1}(r_2)\rho_{R_1 T}(t,r_1)\big)\label{380}.\eea
Using the relations (\ref{370}) and (\ref{380}), we can obtain
\bea P_{RT}(r_2,r_1/t) = P_{R_2R_1}(r_2/t,r_1)P_{R_1 T}(r_1/t) = \mathrm{Trace}\big(\sigma_{RS}(r_1,r_2)\rho_{ST}(t)\big)\label{390}.\eea
where in the second form
\bea \sigma_{RS}(r_1,r_2) = \sqrt{\sigma_{R_1S}(r_1)}\sigma_{R_2R_1}(r_2)\sqrt{\sigma_{R_1S}(r_1)}\label{400}\eea
is the measurement operator corresponding to the two sequential measurements. If the either
measurement is complete, then the combined measurement is also complete.

To take a simple example, suppose that the TPO is a SHO, that the first measurement is
an incomplete measurement of position as given by Eq. (\ref{340}), and that the second is an exact
measurement of momentum, given by $\sigma_{R2R1}(pR) = |p_R\rip\lip p_R|$. With a little work, we can show
that this combined measurement is equivalent to one using squeezed states as measurement
states. The measurement operator for the combined measurement, according to the relation
(\ref{400}) is
\bea \sigma_{RS}(q_R,p_R) = \sqrt{\frac \eta\pi}\left(\exp\left(-\frac\eta 2(q_R-\mathbf q)^2\right)|p_R\rip\lip p_R|\exp\left(-\frac\eta 2(q_R-\mathbf q)^2\right)\right).\label{410}\eea
Matrix elements of this measurement operator in the position representation are
\bea \lip q|\sigma_{RS}(q_R,p_R)|q'\rip = \frac{1}{2\pi}\sqrt{\frac \eta\pi}\exp\left(-\frac\eta 2(q_R- q)^2-\frac\eta 2(q_R-q')^2+ip_R(q-q')\right),\label{420}\eea
where we have used $\lip q|p\rip = (1/\sqrt{2\pi})\exp(ipq)$. In comparison, the position representation of
aa squeezed ground state is
\bea \lip q|0,\eta\rip = \left(\frac\eta\pi\right)^{1/4}\exp\left(-\frac\eta 2 q^2\right).\label{430}\eea
We can label a displaced squeezed ground state by $|p_R,q_R,\eta\rip = \mathbf D(p_R, q_R)|0,\eta\rip$. The position
representation of a displaced squeezed ground state is
\bea \lip q|p_R,q_R,\eta\rip = \left(\frac\eta\pi\right)^{1/4}\exp\left(-\frac\eta 2(q_R-q)^2 + ip_Rq - \frac i 2 p_R q_R\right).\label{440}\eea
Comparing this result with Eq. (\ref{420}), it is apparent that the measurement operator $\sigma_{RS}(q_R,p_R)$
of Eq. (\ref{410}) can be written as
\bea \sigma_{RS}(q_R,p_R) = \frac {1}{2\pi}|p_R,q_R,\eta\rip\lip p_R,q_R,\eta|.\label{450}\eea
We can see the uncertainty principle at work here. The initial inexact position measurement
causes some uncertainty in the momentum, so that even though the subsequent measurement
of momentum is exact, it does not result in an exact value for the momentum of the incoming
SHO.

\section*{Two level system}
The prototype two level system is a spin one-half particle, such as an electron, and the
prototype experiment is the Stern-Gerlach experiment, where spin one-half particles are
deflected by a magnetic field gradient. In principle the transmitter can prepare the spin
to point in any direction, and the receiver can ask in what direction the spin is pointing.
The mystery is that no matter in what direction the transmitter prepares the spin, nor in
what direction the receiver looks for the spin, the measured spin value turns out to be either
plus or minus one-half. The only variable is the probability of these two possible results.
This mystery is not trivial, because the spin carries with it both angular momentum and a
magnetic field. In spite of much effort, this rule has not yet been denied by any experiment
nor has it been explained by any theory other than quantum mechanics. It has important
relevance to the communications problem.

In quantum mechanics, the two states spin up and spin down form a complete orthogonal
set for spin one-half particles. Thus, any spin state can be represented by a two component
vector, and any density operator or measurement operator by a two by two Hermitian matrix.
The identity matrix and the three Pauli spin matrices form a complete set of two by two
Hermitian matrices. We can pursue the spin one-half case by using the following set of
vectors and matrices
\bea |\uparrow\rip = \left(
                       \begin{array}{c}
                         1 \\
                         0 \\
                       \end{array}
                     \right)\hspace{1cm}|\downarrow\rip = \left(
                       \begin{array}{c}
                         0 \\
                         1 \\
                       \end{array}
                     \right)\label{460}
\eea
and
\bea \mathbf I = \left(
                   \begin{array}{cc}
                     1 & 0 \\
                     0 & 1 \\
                   \end{array}
                 \right)\hspace{0.2cm}
                 \mathbf \sigma_x = \left(
                   \begin{array}{cc}
                     0 & 1 \\
                     1 & 0 \\
                   \end{array}
                 \right)\hspace{0.2cm}
                 \mathbf \sigma_y = \left(
                   \begin{array}{cc}
                     0 & -i \\
                     i & 0 \\
                   \end{array}
                 \right)\hspace{0.2cm}
                 \mathbf \sigma_z = \left(
                   \begin{array}{cc}
                     1 & 0 \\
                     0 & -1 \\
                   \end{array}
                 \right).\label{470}\eea
The Pauli spin matrices satisfy the rules
\bea \sigma_i^2 = \mathbf I \hspace{1cm} \sigma_i\sigma_j = -\sigma_j\sigma_i = i\sigma_k,\eea
where $(i,j,k)$ may be any cyclic permutation of $(x,y,z)$. The Pauli spin vector, defined as $\vec\sigma = \sigma_x \hat x + \sigma_y\hat y+\sigma_z\hat z$, where $\hat x$, $\hat y$, and $\hat z$ are the Cartesian coordinate unit vectors, is very
useful in dealing with spin systems. It connects the 2D complex spinor space with the real
3D Cartesian coordinate space of the spin vector model.

In the vector model of angular momentum, a particle with spin $s$ has $2s + 1$ possible states.
The squared length of the spin vector is $s(s + 1)$ and its projection $m_s$ on any chosen axis
takes on values ranging from $s$ to $-s$ separated by unity. Thus a particle of spin 0 is a
singlet. A particle of spin 1/2 has two states, a squared spin vector length of 3/4 and values
of $m_s$ of $(1/2,-1/2)$. A particle of spin 1 has three states, a squared spin vector length of
2, and values of $m_s$ of $(1, 0, -1)$, and so on. Some of these values will appear below in the
discussion of entangled states.

Getting back to the two level spin 1/2 system, we can define the spin operator vector (SOV)
as
\bea \vec s = \frac 1 2\vec\sigma \label{490}.\eea
Note that $\vec s\cdot\vec s = (1/4)(\sigma_x^2+\sigma_y^2+\sigma_z^2) = 3/4\mathbf I$. This represents the squared length of the
SOV. The component of the SOV along a spatial direction $\hat r$ is $\vec s\cdot\hat r$. This operator satisfies
the relation $(\vec s\cdot\hat r)^2 = (1/4)\mathbf I$ and its eigenvalues are therefore $\pm1/2$, the possible values of
$m_s$. Using standard polar coordinates, with the polar axis along $\hat z$, the unit vector $\hat r$ is
\bea \hat r = \sin(\theta) \cos(\phi)\hat x + \sin(\theta)\sin(\phi)\hat y + \cos(\theta)\hat z \label{500}.\eea
The spin state whose component along $\hat r$ is $1/2$ is labeled $|\hat r\rip$. For the sake of convenience,
I will refer to this state as having its spin "pointed" along $\hat r$. It must satisfy the relation
$(\vec s\cdot\hat r)|\hat r\rip = (1/2)|\hat r\rip$. As one can easily verify, this state may be given as
\bea |\hat r\rip = \cos(\theta/2) |\uparrow\rip + \sin(\theta/2)\exp(i\phi)|\downarrow\rip.\label{510}\eea
To go from $\hat r$ to $-\hat r$ involves the transformation $\theta\longrightarrow\pi-\theta$ and $\phi\longrightarrow\pi+\phi$. Thus, the
corresponding spin state pointed along $-\hat r$ is given by
\bea |-\hat r\rip = \sin(\theta/2)|\uparrow\rip -\cos(\theta/2)\exp(i\phi)|\downarrow\rip.\label{520}\eea
These two states are orthogonal and complete, just as are the up and down states. Some
useful relations which describe the properties of the spin 1/2 states are
\bea \lip \hat r|\vec s|\hat r\rip=1/2\hat r\hspace{1cm} |\hat r\rip\lip\hat r| = (\mathbf I+\hat r\cdot\vec\sigma)/2\hspace{1cm}|\lip \hat r|\hat r'\rip|^2 =
(1+\hat r\cdot\hat r')/2\label{530}.\eea

The first of these relations says that the mean value of the SOV for the spin state $|\hat r\rip$ is 1/2
in the direction $\hat r$. The second is consistent with the completeness of the two states $|\hat r\rip$ and
$|-\hat r\rip$. The last is the probability that the receiver by an exact measurement �nds the spin
pointed in the direction $\hat r'$  given that the transmitter prepared it pointed in the direction $\hat r$.
For example, if the transmitter prepares a spin in the $\hat x$ direction, and the receiver measures
the spin component in the $\hat z$ direction, it will find the spin pointing in the $\hat z$ direction half
of the time, and in the $-\hat z$ direction the other half of the time.

To address a more general hypothetical communications problem using spin one-half particles,
we can think of a transmitter which sends out particles whose spins are pointed in various
directions, and a receiver which tries to determine in which of these various directions the
spin was sent. Let us suppose that the sum of the direction vectors $\hat r$ is zero. I do not know
how one might make such a receiver, except for the case of two directions, but the formalism
allows imagining it. For this situation, the density operator representing a spin sent pointing
in the direction $\hat r$ has the form
\bea \rho_{ST}(\hat r) = |\hat r\rip\lip \hat r| = (\mathbf I + \hat r\cdot\vec\sigma) /2\label{540}\eea
and a measurement operator representing the receiver's finding the spin pointing in the
direction $\hat r'$ might have the form
\bea \sigma_{RS}(\hat r') = (2/N)|\hat r'\rip\lip\hat r'| = (\mathbf I+\hat r'\cdot\vec\sigma)/N\label{550},\eea
where the normalization constant $N$ is the number of directions used by the system. Since
the sum of the direction vectors $\hat r$ (or $\hat r'$) is zero, the sum of the measurement operators
is the identity, as required. (We have a bit of a notation problem here, because the Pauli
spin operators and the measurement operators both use the symbol $\sigma$. The context can
distinguish which is meant.)

The probability that the receiver gets the value $\hat r'$ given that the transmitter has sent $\hat r$ is
\bea P(\hat r'/\hat r) = \mathrm{Trace}\big(\sigma_{RS}(\hat r')\rho_{ST}(\hat r)\big) = (1+\hat r'\cdot\hat r)/N\label{560}.\eea
Using Eq. (\ref{560}), one can show that the information transmitted by the system is maximized at
one bit per particle by choosing just two orthogonal states ($N = 2$). This example illustrates again what I believe to be a basic truth of quantum mechanics, namely that the ability to
communicate is fundamentally limited even in the best of circumstances.

In recent times, there has been growing interest in entangled states. The prototype here is
a spin zero particle which decays into two spin one-half particles. Conservation of angular
momentum dictates that the spins of the two resulting particles must be antiparallel. Thus
two separate detectors examining the two particles will have correlated results, no matter
how far apart they are in either space or time, provided that the path does not corrupt the
correlation.

Entanglement involves the case of two spin one-half particles. Call them $A$ and $B$. There
are four possible states of the two. A complete set can be written as
\bea |\uparrow\rip_A|\uparrow\rip_B\hspace{0.6cm}|\uparrow\rip_A|\downarrow\rip_B\hspace{0.6cm}|\downarrow\rip_A|\uparrow\rip_B\hspace{0.6cm}
|\downarrow\rip_A|\downarrow\rip_B\label{570},\eea
or more succinctly as $|\uparrow\uparrow\rip$, $|\uparrow\downarrow\rip$, $|\downarrow\uparrow\rip$, $|\downarrow\downarrow\rip$, where the first arrow symbol refers to the $A$ partricle, and the second arrow symbol refers to the $B$ particle. Entangled states are
mixtures of these elementary states. From the theory of angular momentum, we expect to
find one singlet state with spin zero, and one triplet state with spin one. In this case the SOV is
$\vec s = (1/2) (\vec\sigma_A + \vec\sigma_B) $, where $\vec\sigma_A$ and $\vec\sigma_B$ commute since they refer to di�erent particles.
The squared length of the SOV is given by $\vec s\cdot\vec s = (1/2)(3I+\vec\sigma_A\cdot\vec\sigma_B)$, and its component
along the $\hat z$ axis is $\vec s\cdot\hat z = (1/2) (\sigma_{zA}+\sigma_{zB})$. The states $|\uparrow\uparrow\rip$ and $|\downarrow\downarrow\rip$ belong to the triplet state. They are eigenstates of $\vec s\cdot\vec s$ and of $\vec s\cdot\hat z$. Thus
\bea \vec\sigma_A\cdot\vec\sigma_B |\uparrow\uparrow\rip &=& |\uparrow\uparrow\rip \hspace{0.6cm} \vec\sigma_A\cdot\vec\sigma_B |\downarrow\downarrow\rip = |\downarrow\downarrow\rip\nonumber\\
\vec s\cdot\hat z |\uparrow\uparrow\rip &=& |\uparrow\uparrow\rip \hspace{0.6cm}
\vec s\cdot\hat z|\downarrow\downarrow\rip  = -|\downarrow\downarrow\rip, \label{580}\eea
so that the eigenvalue of $\vec s\cdot\vec s$ in both cases is 2, as appropriate for a particle with spin 1.
The other two angular momentum states are entangled mixtures of the elementary states
$|\uparrow\downarrow\rip$ and $|\downarrow\uparrow\rip$, all of which have a zero value of $\vec s\cdot\hat z$. The state with spin one is $|\psi_+\rip = (1/\sqrt{2})(|\uparrow\downarrow\rip+|\downarrow\uparrow\rip)$.
The state with spin zero is $|\psi_+\rip = (1/\sqrt{2})(|\uparrow\downarrow\rip-|\downarrow\uparrow\rip)$. These states
are also eigenstates of the operator $\vec s\cdot\vec s$, with
\bea \vec\sigma_A\cdot\vec\sigma_B|\psi_+\rip=|\psi_+\rip\hspace{1cm} \vec\sigma_A\cdot\vec\sigma_B|\psi_-\rip=-3|\psi_-\rip\label{590}\eea
so that the eigenvalues of $\vec s\cdot\vec s$ are respectively 2 and 0.

In another context, these states are familiar from the theory of superradiance. If the radiation
field can produce a transition from an up state to a down state, and the two particles are
separated in space by much less than a wavelength, the transitions go from the state $|\uparrow\uparrow\rip$ to
$|\psi_+\rip$ and from $|\psi_+\rip$ to $|\downarrow\downarrow\rip$, and vice versa. The state $|\psi_-\rip$ is decoupled from the radiation field.

In the context of entangled states, one can look for eigenstates of the correlation operators
$\sigma_{iA}\sigma_{iB}$, where $i=x,$ $y,$ and $z$. As it turns out, these are the entangled states
$|\psi_-\rip$ and $|\psi_+\rip$, along with two other entangled states $|\psi_S\rip = (1/\sqrt{2})(|\uparrow\uparrow\rip+|\downarrow\downarrow\rip)$ and $|\psi_D\rip =(1/\sqrt{2})(|\uparrow\uparrow\rip-|\downarrow\downarrow\rip)$. One may verify the relations
\bea \sigma_{xA}\sigma_{xB}|\psi_-\rip &=& -|\psi_-\rip \hspace{0.3cm}\sigma_{yA}\sigma_{yB}|\psi_-\rip = -|\psi_-\rip \hspace{0.3cm}\sigma_{zA}\sigma_{zB}|\psi_-\rip = -|\psi_-\rip \nonumber\\
\sigma_{xA}\sigma_{xB}|\psi_+\rip &=& +|\psi_+\rip \hspace{0.3cm}\sigma_{yA}\sigma_{yB}|\psi_+\rip = +|\psi_+\rip \hspace{0.3cm}\sigma_{zA}\sigma_{zB}|\psi_+\rip = -|\psi_+\rip \nonumber\\
\sigma_{xA}\sigma_{xB}|\psi_S\rip &=& +|\psi_S\rip \hspace{0.3cm}\sigma_{yA}\sigma_{yB}|\psi_S\rip = -|\psi_S\rip \hspace{0.3cm}\sigma_{zA}\sigma_{zB}|\psi_S\rip = +|\psi_S\rip \nonumber\\
\sigma_{xA}\sigma_{xB}|\psi_D\rip &=& -|\psi_D\rip \hspace{0.3cm}\sigma_{yA}\sigma_{yB}|\psi_D\rip = +|\psi_D\rip \hspace{0.3cm}\sigma_{zA}\sigma_{zB}|\psi_D\rip = +|\psi_D\rip.
\nonumber\\\label{600}\eea
It may be seen that the three entangled states having total spin one have positively correlated
spins in two of the three directions and negatively correlated spins in the third direction,
while the spin zero state has negatively correlated spins in all three directions, and in fact
that is so for any direction. Thus it would seem that the spin zero state is the one most useful
for applications of these correlations. Since any two states $\hat r\rip$ and $|-\hat r\rip$ form a complete set
for one spin $1/2$ particle, the four states $|\hat r\hat r'\rip$, $|-\hat r\hat r'\rip$, $|\hat r-\hat r'\rip$,
$|-\hat r-\hat r'\rip$, form a complete set for two spin 1/2 particles. One can show that
\bea |\lip\hat r\hat r'|\psi_-\rip|^2 = (1/4)(1-\hat r\cdot\hat r')\label{610}.\eea
Thus if one receiver looks for one of the particles in directions $\hat r$ or $-\hat r$ and the other looks for
the other particle in directions $\hat r'$ or $-\hat r'$, the probability is $(1/2)(1-\hat r\cdot\hat r')$) if both directions have the same sign, and $(1/2)(1+\hat r\cdot\hat r')$ if they have opposite signs. The probability is zero
for parallel spins, and unity for antiparallel spins. There is no communication involved here,
since neither receiver plays the part of a transmitter. However, such a system has been used
to set up a key for secure communications.

Much has been made of this result as showing that a classical theory is impossible, even
postulating hidden variables. Perhaps this is a question of kicking a dead horse. The wave-particle duality required by quantum mechanics, and demonstrated by experiment, is proof
enough. Such things as Plank's law of black-body radiation and the details of photoemission
defy classical explanations, if I am not mistaken.

\section*{Summary and discussion}
I have discussed measurements in the context of a communications system, and have promoted the idea that measurement operators are related to measurement results in essentially
the same way that density operators are related to transmitter settings. Together they guarantee satisfaction of the uncertainty principle with respect to both state preparation and
state measurement. Their overlap, namely the trace of the product of a density operator
and a measurement operator, gives the probability of the receiver's reading given the transmitter setting, which is the value basic to the communications process. To deal with the
essence of the problem, the discussion was centered on simple harmonic oscillators (SHOs)
and two-level systems (TLSs) as the carriers of information from transmitter to receiver.
In the case of the SHO, the Wigner distributions were shown to be isomorphic with the
quantum density and measurement operators, and also to give a classical picture of the communications process in the common case of a coherent signal plus additive Gaussian noise.
The two-level system has no easy classical counterpart. It is the workhorse of the so-far
successful efforts to show that no classical picture can account for the experimental results.

Looking forward, are there any practical results of this business, aside from the knowledge
that quantum mechanics is necessary to describe the results of some experiments in communication? One, I believe, is that the Wigner picture best describes the business of most
optical (photonic) communications. Transmission of an optical field with bandwidth B Hertz
in a single mode fiber is equivalent to the transmission of SHOs at a rate of B per second.
The Wigner picture therefore treats the zero-point noise field of $h\nu/2$ spectral power (power
per unit bandwidth) as an integral part of the a classical noise field in the fiber. Lossy pieces
of fiber absorb whatever field is traversing the fiber (as in the above discussion of squeezed
fields), and create and maintain the zero-point field through the combined processes of noise
generation and absorption. Coherent amplifiers add their contributions to the noise field,
but even in the case of a dark input (an input with no photons), a coherent amplifier sees
and amplifies the zero-point field, just as it does any other incident field. This reduces the
spontaneous emission noise the amplifier makes on its own, by a factor of two in the case
of pure amplification with no accompanying sources of loss. One further point should be
mentioned, although it is not considered above. In the case of weak non-linearity, such as
exists in a glass fiber, the Wigner picture gives a classical account of the field propagation
to second order in the size of the quantum. And because the Wigner distributions are isomorphic with the quantum density operators, it is always possible in principle to revert to
the quantum picture if that is wanted.

\setcounter{equation}{0}
\numberwithin{equation}{section}
\newpage
\begin{appendix}
\section{A brief review of the quantum theory of a simple harmonic oscillator \label{App1}}
The fundamentals of the quantum mechanics of the simple harmonic oscillator (SHO) are
reviewed here. For simplicity I have scaled things so that $\hbar = 1$ and $\omega_0 = 1$, where $\omega_0$ is the
natural resonance frequency of the oscillator. This scaling makes $\hbar$ the unit of action, and $\omega_0$
the unit of frequency. The position and momentum of the SHO are represented respectively
by the observables $\mathbf q$ and $\mathbf p$. The Hamiltonian of the SHO is
\bea \mathbf H =(\mathbf p^2 + \mathbf q^2)/2\label{A1}.\eea
Both $\mathbf p$ and $\mathbf q$ have dimensions of the square root of energy. The energy quantum $\hbar\omega_0$ is
the energy unit, since $\hbar\omega_0=1$. In both classical and quantum mechanics, the motion of
position and momentum are given by the linear relations $\partial \mathbf q/\partial t = \mathbf p$ and $\partial \mathbf p/\partial t =-\mathbf q$. The
crucial axiom of quantum mechanics is that $\mathbf q$ and $\mathbf p$ are operators that do not commute,
but rather satisfy the commutator relation
\bea \label{A2} [\mathbf q,\mathbf p] \equiv \mathbf{qp}-\mathbf{pq} = i.\eea
This axiom leads directly to the ladder of energy states of the SHO, and satisfies the uncertainty principle's requirement that definite values of $\mathbf q$ and $\mathbf p$ cannot be simultaneously
prescribed. It is customary to define the complex operator $\mathbf a$ and its Hermitian conjugate
$\mathbf a^\dagger$ as
\bea \mathbf a = (1/\sqrt{2})(\mathbf q+i\mathbf p)\hspace{1cm}\mathbf a^\dagger = (1/\sqrt{2})(\mathbf q-i\mathbf p)\label{A3}.\eea
From equations (\ref{A1}) -- (\ref{A3}), one finds the commutator $[\mathbf{a,a}^\dagger] = 1$, and that $\mathbf H = \mathbf a^\dagger\mathbf a + 1/2.$
On the assumption that the SHO has a state $|n\rip$ that is an eigenstate of the operator $\mathbf a^\dagger\mathbf a$
with eigenvalue $n$, that is
\bea \mathbf a^\dagger\mathbf a|n\rip = n|n\rip,\label{A4}\eea
one finds that $\mathbf a|n\rip$ is an eigenstate of $\mathbf a^\dagger\mathbf a$ with eigenvalue $n-1$. (hint: multiply equation (\ref{A4}) by $\mathbf a$ and apply the appropriate commutation rule). Normalization plus the requirement
that the ladder of states stops at $n = 0$ establishes that $n$ must be an integer, and that
\bea \mathbf a|n\rip = \sqrt{n}|n-1\rip\hspace{1cm}\mathbf a^\dagger |n\rip = \sqrt{n+1}|n+1\rip\label{A5}.\eea
Thus one finds the set of energy states, separated by the energy quantum, with the lowest
state $1/2$ quantum above the classical zero.

In the position representation, in view of the commutator (\ref{A2}), the operator $\mathbf p$ becomes
$-i\partial/\partial\mathbf q$, so that the relation $\mathbf a|0\rip = 0$ gives
\bea \lip q|\mathbf a|0\rip = (1/\sqrt{2})(q+\partial /\partial q)\lip q|0\rip = 0. \label{A6}\eea
The normalized solution of this equation is
\bea\label{A7} \lip q|0\rip = \pi^{-1/4} \exp(-q^2/2).\eea
Similarly, in the momentum representation, with $\mathbf q$ becoming $i\partial /\partial p$, one finds
\bea \label{A8} \lip p|0\rip = \pi^{-1/4} \exp(-p^2/2).\eea
Here we see the uncertainty principle at work. Even in its ground (vacuum) state, there is an
uncertainty in both the position and momentum of the SHO. There is no way of preparing the
SHO with a smaller product of uncertainties. It is one of the oddities of quantum mechanics,
that the position and momentum observables of the SHO have a minimum of Gaussian noise
associated with them, and yet the energy shows a set of well defined values.

The coherent states of the SHO are important to our discussion. They are versions of the
ground state displaced to various locations in the phase space of the SHO. There is an unitary
operator that performs such displacements. It is
\bea \mathbf D(p',q') = \exp(ip'\mathbf q - iq'\mathbf p)\hspace{0.5cm}or\hspace{0.5cm} \mathbf D(\alpha) =\exp(\alpha\mathbf a^\dagger - \alpha^*\mathbf a),\label{A9}\eea
where $\alpha = (1/\sqrt{2})(q'+ip')$. I have used primes here to emphasize values rather than
operators. To deal with this operator we need the Baker-Hausdorf (BH) theorem, which
teaches that given two operators, say $\mathbf A$ and $\mathbf B$, whose commutator commutes with each of
them, one has the relations
\bea \exp(\mathbf{A+B}) &=& \exp(\mathbf A)\exp(\mathbf B)\exp(-[\mathbf{A,B}]/2)\nonumber\\
&=&\exp(\mathbf A/2)\exp(\mathbf B)\exp(\mathbf A/2).\label{A10}\eea
Equation (\ref{A2}) generalizes to yield the commutators
\bea [\mathbf q,f(\mathbf p)] = i\partial f/\partial\mathbf p\hspace{0.5cm}or\hspace{0.5cm}
[\mathbf p,f(\mathbf q)] = -i\partial f/\partial\mathbf q.\label{A11}\eea
Using equation (\ref{A11}) and the BH theorem, one can show that $\mathbf D(p',q')\mathbf q\mathbf D^\dagger(p',q')=\mathbf q-q'$ and that $\mathbf D(p',q')\mathbf p\mathbf D^\dagger(p',q')=\mathbf p-p'$. The coherent states have the form
\bea |\alpha\rip = \mathbf D(\alpha) |0\rip =\exp\left(-\frac{1}{2}\alpha\alpha^*\right)\exp(\alpha\mathbf a^\dagger)|0\rip\label{A12},\eea
where Eq. (\ref{A9}), the BH theorem, and $\mathbf a|0\rip = 0$ are used. The number representatives of
the coherent states are
\bea \lip n|\alpha\rip = \exp\left(-\frac 1 2\alpha\alpha^*\right)\frac{\alpha^n}{\sqrt{n!}}.\label{A13}\eea
The position and momentum representatives of the coherent states are
\bea \lip q|\alpha\rip = \pi^{-1/4}\exp\left(-i\frac{1}{2}q'p'\right)\exp(ip'q)\exp\left(-\frac 1 2(q-q')^2\right)\nonumber\\
\lip p|\alpha\rip = \pi^{-1/4}\exp\left(i\frac{1}{2}q'p'\right)\exp(-iq'p)\exp\left(-\frac 1 2(p-p')^2\right),\label{A14}\eea
where, as above, $\alpha=(1/\sqrt{2})(q' + ip')$. Pertinent to the discussion of measurement is that
the eigenstates of number, position, or momentum form complete orthogonal sets. That is,
one has
\bea \lip n|n'\rip = \delta(n,n')\hspace{0.6cm}\lip q|q'\rip = \delta(q-q')\hspace{0.6cm}\lip p|p'\rip = \delta(p-p')\nonumber\\
\sum|n\rip\lip n| = \mathbf I\hspace{0.6cm}\int\df q|q\rip\lip q| = \mathbf I\hspace{0.6cm}\int\df p|p\rip\lip p| = \mathbf I,\label{A15}\eea
where $\mathbf I$ is the identity operator, and where the sum and integrals cover the complete range
of their arguments. The coherent states are not orthogonal, but they are complete. They
are called overcomplete. One can show that if $|\alpha\rip$ and $|\beta\rip$ represent two coherent states
\bea |\lip\beta|\alpha|^2 = \exp(|\alpha-\beta|^2)\hspace{1cm}\frac 1\pi \int\int\df^{(2)}\alpha|\alpha\rip\lip\alpha|=\mathbf I,\label{A16}\eea
where $\df^{(2)}\alpha$ is the area differential in the complex $\alpha$ plane.
\setcounter{equation}{0}
\newpage
\section{Density matrices and Wigner distributions \label{App2}}
The basic properties of the Wigner distributions and their cousins the coherent state distributions are reviewed here. The Wigner distributions are real functions in the phase space
of the simple harmonic oscillator (SHO) that are isomorphic with the density operators of
quantum mechanics. In many cases they can be used to visualize measurement processes.
The coherent state distributions are useful in the analysis of most situations involving signals
and Gaussian noise.

The Wigner distribution can be defined in terms of a characteristic function
\bea \int\df p\int\df q W(p,q)\exp(ip'q-iq'p) = \mathrm{Trace}\big(\mathbf\rho\exp(ip'\mathbf q-iq'\mathbf p)\big).\label{B1}\eea
Equation (\ref{B1}) gives a quantal-classical correspondence. Any such correspondence requires a particular ordering of the non-commuting quantum operators. In the case of the Wigner density,
the ordering is called symmetrical, since the ordering of the p and q terms in the exponential operator is unimportant. Eq. (\ref{B1}) is expressed so that the exponential operator is the displacement operator $\mathbf D(p',q')$ discussed in Appendix \ref{App1}. Thus, the right side of Eq. (\ref{B1}) can
be written simply as $\lip D(p', q')\rip$, where the angular brackets indicate a mean value. (The
mean value of any operator $\mathbf O$ is given by $\mathrm{Trace}(\mathbf O)$ and is more simply written as $\lip\mathbf O\rip$.)

The Baker-Hausdorf theorem, discussed in Appendix \ref{App1}, shows that
\bea \mathbf D(p',q') = \exp(ip'\mathbf q - iq'\mathbf p) = \exp(-iq'\mathbf p/2)\exp(ip'\mathbf q)\exp(-iq'\mathbf p/2).\label{B2}\eea
Within the trace operation, operators may be cyclically permuted. Expanding $\lip D(p', q')\rip$
from Eq. (\ref{B1}) in the position representation yields
\bea\lip D(p', q')\rip&=&\int\df q\lip q|\exp(-iq'\mathbf p/2)\mathbf\rho\exp(-iq'\mathbf p/2)\exp(ip' q)\nonumber\\
&=&\int\df q\lip q-\frac 1 2q'|\mathbf\rho|q+\frac 1 2q'\rip\exp(ip' q).\label{B3}\eea
Equation (\ref{B3}) allows one to extract from equation (\ref{B1}) the relation
\bea \lip q-\frac 1 2q'|\mathbf\rho|q+\frac 1 2q'\rip = \int\df pW(p,q)\exp(-iq'p)\label{B4},\eea
whence, by Fourier transform, we get
\bea W(p,q) = \frac{1}{2\pi}\int\df q'\lip q-\frac 1 2q'|\rho|q+\frac 1 2q'\rip\exp(iq'p)\label{B5}.\eea
Equations (\ref{B4}) and (\ref{B5}) demonstrate the isomorphism of the Wigner distributions with the
corresponding density operators. Note that if one sets $q'= 0$ in Eq. (\label{B4}), it shows that the
integral over $p$ of the Wigner distribution gives the probability distribution of the position
$q$. Similarly, the integral over $q$ of the Wigner distribution gives the probability distribution
of the momentum $p$.

Of importance to our theme is the relation
\bea \int\df p\int\df q W_1(p,q)W_2(p,q) = (2\pi)^{-1}\mathrm{Trace}(\mathbf\rho_1\mathbf\rho_2),\label{B6}\eea
where $W_1(p,q)$ and $W_2(p,q)$ are the Wigner distributions corresponding respectively to the
density operators $\mathbf\rho_1$ and $\mathbf\rho_2$. The relation (\ref{B6}) can be easily demonstrated using equation
(\ref{B5}). As discussed in the main text, Wigner densities can emulate measurement operators as
well as density operators, and Eq.(\ref{B6}) allows the crucial conditional probability $P(r/t)$ of the
communication process to be given by the overlap of two Wigner densities, one related to
the transmitter setting $t$ and the portion of the path from the transmitter to the location $S$,
the other related to the receiver reading $r$ and the portion of the path from $S$ to the receiver.

Another sometimes valuable distribution function is the coherent state distribution. It is
often called simply the $P$ distribution. Its definition comes from the expansion
\bea \rho = \int\int\df^{(2)}(\alpha)P(\alpha)|\alpha\rip\lip\alpha|,\label{B7}\eea
where $\df^{(2)}(\alpha)$ is the differential area in the complex $\alpha$ plane and $P(\alpha)$ is a real distribution
function of the real and imaginary parts of $\alpha$. A characteristic function for the coherent
state distribution similar to that for the Wigner distribution is
\bea\int\int\df^{(2)}(\alpha)P(\alpha)\exp(\lambda\alpha^*-\lambda^*\alpha) = \mathrm{Trace}\big(\mathbf\rho\exp(\lambda\mathbf a^\dagger)\exp(-\lambda^*\mathbf a)\big),\label{B8}\eea
where $\lambda$ is a complex expansion parameter. This relation can be veri�ed using Eq. (\ref{B7}).
Note that the operators in the trace expression are in normal order for the coherent state distribution, while they are in symmetrical order for the Wigner distribution. Normal order
means that all factors of $\mathbf a^\dagger$ are to the left of all factors of $\mathbf a$. A difficulty with the coherent state distribution is that, unlike the Wigner distribution, it cannot be easily inverted to find
matrix elements of the density operator.

The Wigner distribution and the coherent state distribution are intimately related. If we
insert equation (\ref{B7}) into equation (\ref{B5}) and use equation (\ref{A14}) of Appendix A, the result is
\bea W(p,q) = \pi^{-1}\int\df p'\int\df q' P(p',q')\exp\big(-(p-p')^2 - (q-q')^2\big)\label{B9},\eea
where as before, we have used $\alpha = (q'+ip')/\sqrt{2}$, and $\df q'\df p' P(p',q') = \df^{(2)}(\alpha)P(\alpha)$.
Note that I am using the symbol $P$ generically, meaning 'the probability distribution of'. Thus
the Wigner distribution is a Gaussian convolution of the coherent state distribution, in effect
adding to it one half quantum of Gaussian noise. One can show this result also from the
characteristic functions by using the B-H theorem.

An important class of distributions related to the communications problem are those involving signal plus Gaussian noise (as in thermal noise). The density operator corresponding to
this situation can be written in the form
\bea\mathbf\rho = \big(1-\exp(-\eta)\big)\exp\big(-\eta(\mathbf a^\dagger -\alpha^*)(\mathbf a-\alpha)\big),\label{B10}\eea
where $\eta = \ln(1+\bar n^{-1})$ in which $\bar n$ is the mean number of noise quanta, excluding the zero-
point half quantum, and $\alpha$ represents the signal. The coherent state expansion of the same
density operator is
\bea \mathbf\rho = \frac{1}{\pi\bar n}\int\int\df^{(2)}(\beta)\exp\left(-\frac{|\beta-\alpha|^2}{\bar n}\right)|\beta\rip\lip\beta|,\label{B11}\eea
where $|\beta\rip$ is a coherent state. Comparing Eq.(\ref{B7}), we find that in Eq. (\ref{B11}),
\bea \frac{1}{\pi \bar n}\exp\left(-\frac{|\beta-\alpha|^2}{\bar n}\right)=P(\beta).\label{B12}\eea
Finally, the Wigner distribution is given by
\bea W(p,q/p',q') = \frac{1}{\pi(2\bar n+1)}\exp\left(-\frac{(q-q')^2+(p-p')^2}{2\bar n+1}\right),\label{B13}\eea
where, as before, $\alpha= (q' + ip')/\sqrt{2}$.
The equivalence of Eqs (\ref{B10}) and (\ref{B11}) can be
shown without loss of generality by taking $\alpha = 0$ and taking matrix elements in the energy
representation. Equation (\ref{B13}) can be derived from equation (\ref{B11}) using equation (\ref{B9}). The
Wigner distribution, as we have noted before, adds the one-half quantum of zero-point noise
to the total noise energy.

\end{appendix}

\begin{thebibliography}{99}

\bibitem{G61} J. P. Gordon, ``Information capacity of a communications channel in the presence of quantum effects," in Advances in Quantum Electronics, J. R. Singer ed. Columbia University Press 1961
\bibitem{G62} J. P. Gordon, ``Quantum effects in communications systems," Proceedings of the IRE pp. 1898 -- 1908, (1962)
\bibitem{G64} J. P. Gordon, ``Noise at optical frequencies; Information theory," Fermi conference, Varenna (1963). Published in Proc. Int. School Phys. Enrico Fermi, Course XXXI, pp. 156 -- 181 , (1964)
\bibitem{G66} J. P. Gordon and W. H. Louisell  ``Simultaneous measurements of non-commuting observables," Physics of Quantum Electronics, P. L. Kelley, M. Lax, and P. E. Tannenwald, Eds. New York: McGraw-Hill, 1966, pp. 833 -- 840 (1966)
\bibitem{SympCleo14} Special Symposium in Memory of James P. Gordon, CLEO, San Jose 2014, \begin{verbatim}http://www.osa.org/en-us/foundation/donate/current_campaigns/
    james_p_gordon_memorial_speakership/gordon_symposium/\end{verbatim}
\bibitem{MarkCleo14} M. Shtaif, ``The birth of quantum communications," slides and video, CLEO 2014, available on the website of the Optical Society of America (OSA)
\begin{verbatim}http://www.osa.org/en-us/foundation/donate/current_campaigns/
    james_p_gordon_memorial_speakership/gordon_symposium/
    gordon_symposium_videos_(4)/\end{verbatim}

\end{thebibliography}
\end{document}